# Unusual Magnetic Fields of Uranus and Neptune: Metallic Fluid Hydrogen

W. J. Nellis

Department of Physics, Harvard University
Cambridge, MA 02138, USA
nellis@physics.harvard.edu

I. Introduction

NASA's Voyager 2 spacecraft discovered the unusual non-dipolar and non-axisymmetric magnetic fields of Uranus and Neptune (U/N) in the 1980's [1]. Those magnetic fields are made primarily by degenerate, metallic, fluid H (MFH) [2,3]. The magnetic fields of U/N are both ~$2 \times 10^{-5}$ Tesla [4], comparable to that of Earth. U/N are called the Ice Giant Planets because of their likely chemical compositions. If the magnetic fields of U/N are force-fit to effective dipolar fields, the effective magnetic axes are tilted $59^0$ and $47^0$, respectively, from their respective rotational axes and the effective dipole centers are offset by ~0.33 $R_U$ and ~0.55 $R_N$, respectively, from the centers of U/N, where $R_U$ and $R_N$ are the radii of U/N. In contrast, the magnetic fields of Earth, Saturn, Jupiter and its moon Ganymede are dipolar with their dipolar axes aligned within a few degrees of their respective rotational axes.

The cause of the enormous difference between planetary dipolar fields and non-dipolar/non-axisymmetric magnetic fields has been a major scientific question since the discovery of the magnetic fields of U/N. Planetary magnetic fields are made by dynamo action: convection of electrically conducting fluids at high pressures P and temperatures T in planetary interiors [5]. The cause of the unusual magnetic fields of U/N lies in measured physical properties of fluids that generate those magnetic fields.

Dynamic compression with a two-stage light-gas gun achieves pressures up to ~300 GPa ($3 \times 10^6$ bar=3 Mbar) and temperatures up to several 1000 K in fluids representative of the outer regions of U/N in which unusual magnetic fields are made. Pressures and temperatures range up to ~700 GPa and 7000 K near the centers of U/N [6,7]. A substantial body of experimental data has been measured both by Voyager 2 and on Earth under dynamic compression for fluids expected in the Ice Giants at representative extreme conditions. The prime goal of this paper is to review this experimental database in the context of what those measured results suggest about the likely nature of the deep interiors of Uranus and Neptune, particularly with respect to dense fluid H. Ices and other issues are discussed elsewhere [3].

II. Chemical composition

U/N have similar sizes, densities, magnetic and gravitational fields and probably chemical compositions, although exact chemical compositions of U/N are unknown. Uranus and Neptune are probably composed of three types of materials [8-10]: (i) Gas, an H-rich mixture of H and He, (ii) Ice, a mixture primarily of $H_2O$, $CH_4$, and $NH_3$, and (iii) Rock, a mixture of silicate oxides and Fe/Ni. H and He have solar chemical abundances of about 91 at.% and 8 at.%, respectively [11]. Planetary "ices" are hydrogenous compounds with relatively abundant reactive elements O, N, and C. U/N have masses 14.5 and 17.2 $M_E$, respectively, where $M_E$ is the mass of Earth. Total mass of the Giant Planets Jupiter, Saturn, Uranus and Neptune is 445 $M_E$. Total mass of the Rocky Planets Earth, Venus, Mars and Mercury is 2.0 $M_E$. Most of the mass of the solar system is hydrogen and most of that is metallic fluid H at high internal P and T in the Giant Planets.

Nebular "ices" and $H_2$ accreted to form U/N. Accreted molecular species most probably decomposed and re-reacted at extreme interior pressures and temperatures. The probable reason little nebular "ice" exists in the Ice Giants is based on measured electrical conductivities, pressure-density equation-of-state data, emission temperatures, and pulsed Raman spectroscopy of a large number of small molecular fluids under dynamic compression. Those experiments began in ~1980 when Voyager 2 departed Saturn and headed for Uranus [12,13].

III. Planetary fluids: dynamic compression experiments

The vast majority of measured properties of fluids at planetary P/Ts have been made under dynamic compression up to few 100 GPa (1 Mbar) and few 1000 K. MFH is "cold" degenerate condensed matter with $T/T_F \ll 1$ where $T_F$ is Fermi temperature. Electrical conductivity of MFH is ~2000/(Ω-cm) measured at 140 GPa (1.4 Mbar), 0.64 mol H/cm$^3$ (9-fold compressed H density in liquid $H_2$), and ~3000 K [14-17]. Temperatures of 2000-3000 K are expected to inhibit formation of intermolecular H bonds. The measured density of metallization of MFH is essentially the same metallization density predicted by Wigner and Huntington in 1935, 0.62 mol H/cm$^3$ [18]. MFH was made experimentally *in situ* under quasi-isentropic multiple-shock compression achieved by reverberating a shock wave in liquid $H_2$ contained between two anvils. Magnetic fields of U/N are probably made by dynamo convection of a fluid composed mostly of MFH. Experiments were also performed under single-shock compression. Multiple-shock compression achieves relatively high pressures and relatively low temperatures, whereas single-shock compression achieves relatively low pressures and relatively high temperatures [19].

MFH was achieved by impact onto a sample holder containing liquid $H_2$ at 20 K. The impactor was accelerated to velocities as high as 7 km/s by a two-stage light-gas gun. The liquid-$H_2$ sample was 25 mm in diameter and 0.5 mm thick. Impactor mass was 20 g with maximum kinetic energy ~0.5 MJ, comparable to total energy of $10^{12}$ protons and antiprotons at Fermi Laboratory. Similar experimental configurations were used to measure properties of a large number of potential planetary materials. Typical experimental lifetimes were ~100 ns, sufficiently long to achieve thermal equilibrium in dense fluid H and sufficiently brief to inhibit diffusion of T and H atoms out of the sample holder during experimental lifetimes.

Because neither exact compositions nor exact interior pressures and temperatures in the Ice Giants are known, experiments were conducted on an ensemble of likely planetary materials over a wide range of likely pressures and temperatures. Trends in the resulting measurements are used as a basis for speculation on the nature of planetary interiors. Liquids investigated under multiple-shock compression include $H_2/D_2$[16], $O_2$[20], $N_2$[21], $H_2O$[22,23], and Synthetic Uranus (SU) [7]. SU is a single-phase liquid at ambient, which is a representative planetary mixture of polar molecules $H_2O$, $NH_3$, and $C_3H_8O$ (isopropanol). Liquids investigated under single-shock compression include He[24], $H_2/D_2$[25-30], $N_2$[31-34], $O_2$[35,36], CO[37], $CO_2$[38], $CH_4$[37,39,40,44], $NH_3$,[13,40], $H_2O$[13,41,42], $C_6H_6$[43,44,46], $C_4H_8$,[44-46], $CH_2$,[43] and SU[47]. Photographs of the two-stage gun facility are published [48].

IV. Planetary fluids: static compression experiments

Relatively few measurements on planetary materials at representative conditions have been made under static compression in laser-heated diamond-anvil cells (DAC). The main reason is chemical corrosion and sample diffusion out of a DAC at high pressures and temperatures during experimental lifetimes of a few minutes. Spectroscopies have been measured on $H_2O$, $NH_3$, $CH_4$ and $H_2$ at pressures ranging from 20 to 120 GPa and temperatures as high as 2500 K [49-53].

V. Interior models of Uranus and Neptune

Voyager 2 measured gravitational fields of U/N from which gravitational moments were derived, from which likely radial mass distributions were derived [54]. Densities increase slowly from zero at $R_U$ and $R_N$ up to ~0.6 g/cm$^3$ as radii decrease from $R_U$ and $R_N$ down to normalized radii of ~90% of $R_U$ and $R_N$. At still smaller radii densities increase dramatically from ~0.6 g/cm$^3$ up to ~4-5 g/cm$^3$ near planetary centers. The outer regions composed of Gas (H-He) are called envelopes; the inner regions composed of Ice/Rock are called cores.

Measured electrical conductivities of dense fluid H start at 93 GPa and reach 2000/(Ω-cm), typical of a strong-scattering metal, at 140 GPa, 0.64 g/cm$^3$ and ~2600 K [16]. Although gravitational moments determined from Voyager 2 data are delocalized radial constraints and a planetary density of 0.6 g/cm$^3$ might be caused by Gas mixed with some Ice/Rock, it is reasonable to expect that at 0.6 g/cm$^3$ in U/N pressure is ~100 GPa, which is near completion of the crossover from semiconducting to poor-metallic fluid H near the bottom of the envelopes and tops of the cores. Thus, it appears that the interfacial region between envelope and core coincides or nearly so with the crossover from nonmetallic to poor-metallic H. Further, at pressures and temperatures above 100 GPa and 2000 K, whichever small molecules constitute ice decompose to an H-rich mixture that might have a substantial fraction of MFH. In short MFH probably exists in substantial amounts in the cores up to normalized radii of 0.9 $R_U$ and $R_N$. Electrical conductivity of MFH is a factor of 100 larger than the value of 20/(Ω-cm) measured for ices prior to Voyager 2 [55]. Dynamo action of MFH probably generates the dominant contributions to the magnetic fields of U/N.

Analysis of Voyager 2 data led to the suggestion that the magnetic fields of U/N are made at radii relatively close to the outer radii of those planets [56,57]. Magnetic fields produced close to the surfaces of U/N can be expected to have multi-polar contributions. That is, a dipolar magnetic field falls off with distance as 1/r$^3$, where r is distance from the center of the dipole. Multi-polar magnetic fields fall off faster than 1/r$^3$, and thus are more likely to be observed by spacecraft if they are made close to the surface of a planet.

VI. Conclusions

The magnetic fields of the Ice Giant Planets Uranus and Neptune (U/N) are unique in the solar system. Based on a substantial database measured on Earth for representative planetary fluids at representative dynamic pressures up to 200 GPa (2 Mbar) and a few 1000 K, the complex magnetic fields of U/N are (i) probably made primarily by degenerate metallic fluid H (MFH) at or near the crossover from the H-He envelopes to "Ice" cores at ~100 GPa (Mbar) pressures and normalized radii of ~90% of the radii of U/N; (ii) because those magnetic fields are made relatively close to the surfaces of U/N, non-dipolar fields can be expected as observed; (iii) the "Ice" cores are most probably a heterogeneous fluid mixture of H, N, O, C, Fe/Ni and silicate-oxide species and their mutual reaction products at high pressures and temperatures, as discussed elsewhere [3].